# Robust memristors based on layered two-dimensional materials

Miao Wang[1†], Songhua Cai[2†], Chen Pan[1], Chenyu Wang[1], Xiaojuan Lian[1], Ye Zhuo[3], Kang Xu[1], Tianjun Cao[1], Xiaoqing Pan[2], Baigeng Wang[1], Shijun Liang[1], J. Joshua Yang[3*], Peng Wang[2*] & Feng Miao[1*]

**Van der Waals heterostructure based on layered two-dimensional (2D) materials offers unprecedented opportunities to create materials with atomic precision by design. By combining superior properties of each component, such heterostructure also provides possible solutions to address various challenges of the electronic devices, especially those with vertical multilayered structures. Here, we report the realization of robust memristors for the first time based on van der Waals heterostructure of fully layered 2D materials (graphene/$MoS_{2-x}O_x$/graphene) and demonstrate a good thermal stability lacking in traditional memristors. Such devices have shown excellent switching performance with endurance up to $10^7$ and a record-high operating temperature up to 340°C. By combining *in situ* high-resolution TEM and STEM studies, we have shown that the $MoS_{2-x}O_x$ switching layer, together with the graphene electrodes and their atomically sharp interfaces, are responsible for the observed thermal stability at elevated temperatures. A well-defined conduction channel and a switching mechanism based on the migration of oxygen ions were also revealed. In addition, the fully layered 2D materials offer a good mechanical flexibility for flexible electronic applications, manifested by our experimental demonstration of a good endurance against over 1000 bending cycles. Our results showcase a general and encouraging pathway toward engineering desired device properties by using 2D van der Waals heterostructures.**

[1]National Laboratory of Solid State Microstructures, School of Physics, Collaborative Innovation Center of Advanced Microstructures, Nanjing University, Nanjing 210093, China. [2]College of Engineering and Applied Sciences and Collaborative Innovation Center of Advanced Microstructures, Nanjing University, Nanjing 210093, China. [3]Department of Electrical and Computer Engineering, University of Massachusetts, Amherst, MA 01003, USA.

† These authors contributed equally to this work. Correspondence and requests for materials should be addressed to F. M. (email: miao@nju.edu.cn), J. J. Yang (email: jjyang@umass.edu) or P. W. (email: wangpeng@nju.edu.cn).

Memristor emerges as a top candidate for future storage and neuromorphic computing technologies[1-13] due to many advantages including device scalability, multi-state switching, fast switching speed, high switching endurance and CMOS compatibility[8,14-24]. While most research and development efforts have been focused on improving device switching performance in optimal conditions, the reliability of memristors in harsh environments such as at high temperature and on bending substrates has been much less studied. Since the programming processes in the memristors based on traditional oxide materials mostly rely on ion moving and ionic valence changing[25-27], the thermal instability at elevated temperatures could result in undesirable device failure.[28,29] Thus, to the best of our knowledge, there has been no reliable switching behaviors observed in memristors at temperatures above 200°C[29,30], which has greatly limited their potential applications in harsh electronics, such as those demanded in aerospace, military, automobile, geothermal, oil and gas industries. While common high temperature electronic materials, such as SiC and III-nitride[31,32], are not adoptable in fabricating memristors, searching for new materials and structures for robust memristors with good performance becomes highly desirable.

Van der Waals (vdW) heterostructures by stacking two-dimensional (2D) layered materials together[33-41] can combine superior properties of each component and offer a unique approach to address such challenge. 2D materials represented by graphene have shown excellent structural stability[42,43] and electrical properties, which may lead to tremendous improvements in the robustness of electronic devices. For instance, graphene possesses unparalleled breaking strength, and ultra-high thermal and chemical stabilities[44]; molybdenum disulfide ($MoS_2$) has shown good flexibility, large Young's modulus (comparable to stainless steel),[45] and excellent thermal stability up to 1100°C[43]; various functionalized 2D material layers or certain grain boundaries have shown switching behaviors.[46-57] Since both the thickness and roughness of 2D layered materials can be controlled accurately at atomic scale, the reliability and uniformity of the electronic devices based on such materials and their vdW heterostructures could also be significantly improved.

Here we report robust memristors based on a vdW heterostructure made of fully layered 2D materials (graphene/$MoS_{2-x}O_x$/graphene), which exhibit repeatable bipolar resistive switching with endurance up to $10^7$ and high thermal stability with operating temperature up to 340°C. $MoS_{2-x}O_x$ layer was found to be responsible for the high thermal stability of the devices by performing high temperature *in situ* high-resolution transmission electron microscopy (HRTEM) studies. Further *in situ* scanning transmission electron microscopy (STEM) investigations on the cross section of a functional device revealed a well-defined conduction channel and a switching mechanism based on the migration of oxygen ions. The atomic layered structure of both memristive material ($MoS_{2-x}O_x$) and electrodes (graphene) was found to be well maintained during the switching processes, which plays a crucial role in determining the robustness of the devices. Finally, the mechanical flexibility of such devices was demonstrated on Polyimide (PI) substrate with a good endurance against mechanical bending of over 1000 times.

**Device fabrication and switching performance.**

Fig. 1a shows the schematic drawing and the crystal structure of graphene/$MoS_{2-x}O_x$/graphene (GMG) devices. Multi-layer graphene (~8 nm thick) and $MoS_2$ (~40 nm thick) membranes were mechanically exfoliated and deposited on $SiO_2$/Si wafers. We oxidized $MoS_2$ membranes at 160°C for 1.5 h in ambient air to obtain the layered $MoS_{2-x}O_x$ membrane, where x ≈ 0.3 according to Energy Dispersive X-Ray Spectroscopy (EDS) analysis (Fig. 5d). Fig. 1b shows the optical image of a typical GMG device and measurement setup. The stacked vdW heterostructure was obtained by using standard polyvinyl alcohol (PVA) transfer method. Figs. 1c and 1d show the cross-section high angle annular dark-field (HAADF) and HRTEM images respectively. Remarkably, the $MoS_{2-x}O_x$ layer exhibits excellent layered crystal structure after the oxidization process. Compared to the previous works of replacing either electrode or switching layer by layered materials to improve performance[46-57], one major advantage of our GMG devices based vdW heterostructure is the realization of the atomically sharp interfaces between the switching layer and the electrodes (as shown in Fig. 1d). This is not achievable in traditional metal/oxide/metal based memristors fabricated by sputtering and evaporating[25,26]. Since the switching performance of memristors is largely affected by the roughness of the switching interface[7,58,59], high performance could be expected in the GMG devices.

We then examined the switching performance of our GMG devices by using four-probe electrical measurements (to rule out the line resistance). Fig. 2a shows the repeatable bipolar switching curves of a typical GMG device. The switching polarity is determined by the electroforming voltage polarity, i.e. positive/negative electroforming resulting in ON switching with positive/negative bias and OFF switching with negative/positive bias (see Supplementary Fig. 1a and Supplementary Note 1 for more details). Higher ON/OFF ratio can be achieved with higher current compliance for electroforming and set process (Supplementary Fig. 1b). The switching voltage for the first set operation (orange curve in Fig. 2a) is only slightly larger than that of the subsequent normal set operation (blue I-V), suggesting that no dramatic electroforming process is required for device operations. As shown in Fig. 2b, over $2 \times 10^7$ switching cycles were observed by applying fixed voltage pulses (1 μs width with +3.5 V for set and -4.8 V for reset). We also used fixed voltage pulses (+3 V for set and -4 V for reset) with progressively broadened width to test the switching speed of our GMG devices. The results show that the GMG device could switch in less than 100 ns and sustain its resistance state under the following wider reading pulses (Fig. 2c). Here we note that this measured switching speed is limited by the parasitic capacitance of the device, not the intrinsic switching speed. In addition, a good retention (~$10^5$ s) of both ON and OFF states at room temperature was shown in Fig. 2d. During all electrical measurements, the voltages were applied on the graphene top electrode and the graphene bottom electrode was always grounded.

In order to investigate the role of the high-quality interface in the GMG devices, we carried out control experiments by fabricating devices based on Au/MoS$_{2-x}$O$_x$ (~40 nm)/Au (AMA) structure (using the same MoS$_{2-x}$O$_x$ membranes as in GMG devices). Repeatable switching behaviors were also observed in AMA devices. We carefully analyzed and compared the statistics of the switching parameters in AMA and GMG devices. Compared with the GMG devices, the AMA devices have shown much boarder distributions in both reset and set voltages (see Supplementary Figures 2a-e), suggesting a larger variance and less reliable resistive switching in AMA devices. A much lower switching endurance with a larger cycle-to-cycle resistance variance was also observed in the AMA devices, as shown in Supplementary Fig. 2f. The cross-section HRTEM studies of an AMA device (see Supplementary Fig. 3) revealed a rough interface between the Au electrode and MoS$_{2-x}$O$_x$ layer, indicating a crucial role possibly played by the high-quality switching interface in memristor devices based on layered materials.

**High-temperature performance and thermal stability.**

To further examine the thermal stability of the GMG devices, we measured their switching performance at elevated temperatures. Various switching curves of a typical GMG device at ambient temperatures ranging from 20°C to 340°C are shown in Fig. 3a. The device remained fully functional at 340°C. We also used voltage pulses (fixed width of 1 μs) to test the switching repeatability of the GMG devices at elevated temperatures. As shown in Fig. 3b, the GMG device exhibits a good cycle to cycle reproducibility and a stable switching window at three chosen temperatures (100°C, 200°C and 300°C). The retention of both ON and OFF states at elevated temperatures was also examined. As shown in Fig. 3c, we observed no significant resistance change of ON/OFF state over an approximately $10^5$ s testing time at 160°C and 340°C respectively, suggesting a great reliability of the GMG devices at elevated temperatures. The demonstrated operating temperature of 340°C is record-high for memristors, and much higher than the previously reported highest temperature of 200°C[30], suggesting potential applications of GMG devices as high-density memory/computing units in future high-temperature harsh electronics.

Compared with the amorphous oxide switching layer used in the traditional memristors, the layered crystal structure of $MoS_{2-x}O_x$ could be responsible for the robustness of GMG devices in terms of high thermal stability. In order to verify this, we performed *in situ* HRTEM studies on the $MoS_{2-x}O_x$ membranes at high temperatures (Fig. 4a-d). Remarkably, the $MoS_{2-x}O_x$ membrane maintains excellent crystal structure at temperatures up to 800°C. Ion migration starts to occur only when the temperature increases to 900°C or higher (see Supplementary Fig. 5). As a matter of fact, pristine $MoS_2$ stays stable at very high temperatures. For example, $MoS_2$ can still be used as a lubricant at 1300°C[60]. Thus, the superior high thermal stability of the $MoS_{2-x}O_x$ layer should be attributed to the structural stability of $MoS_2$ (since the $MoS_{2-x}O_x$ layer was obtained by oxidizing $MoS_2$). Such high thermal stability has not been reported in the amorphous oxides used in traditional memristors (such as in Titanium oxides and Tantalum oxides, with phase transition temperatures at 300°C[61] and 650°C[62] respectively). The thermal stability of $MoS_{2-x}O_x$ layer also enables the non-switching regions in the devices to maintain their original state at elevated temperatures, hence ensures the stable switching of the GMG devices.

***In situ* STEM study and microscopic pictures.**

In order to further reveal the nature of the robustness of GMG devices, it is essential to acquire the microscopic pictures during the electroforming and switching processes. We fabricated electron-transparent cross-sectional TEM samples with a similar GMG structure (Pt/graphene/MoS$_{2-x}$O$_x$/graphene/conductive silicon substrate, see details in methods section), followed by *in situ* STEM experiments. This makes it feasible for real-time observation of any voltage-induced structural change, particularly in the conduction channel region, with typical HAADF-STEM and EDS results of different states shown in Figs. 5a-c (pristine, ON and OFF states respectively). The whole area of the cross-sectional device was investigated during the *in situ* STEM experiment to confirm that there is no region that is critical for the switching except the one shown in Fig. 5. After the electroforming with a positive staircase sweep voltage (from 0 to +4.5 V) applied on the top electrode (Pt), a dark-contrast region in the MoS$_{2-x}$O$_x$ layer appeared (in the ON state, shown in Fig. 5b) and barely changed when the device was switched to the OFF state (Fig. 5c). This observation suggests that a conduction channel was formed with some noticeable composition change, as the intensity of the HAADF image is monotonically proportional to atomic number Z[63]. In addition, EDS line-scan analysis was employed to study the elemental composition variations of the MoS$_{2-x}$O$_x$ layer for three different states as shown in Figs. 5d-f respectively (the scanning directions are indicated by the green arrows in Figs. 5a-c). In the pristine state, the MoS$_{2-x}$O$_x$ layer shows a uniform atom distribution in stoichiometric proportions with an averaged atomic ratio of molybdenum to sulfur and oxygen at approximate 0.5 (Mo : (S+O) ≈ 1 : 2, as shown in Fig. 5d and Supplementary Fig. 6). This suggests that sulfur vacancies in MoS$_{2-x}$O$_x$ could be mostly occupied by oxygen after the thermal oxidation (as schematically shown in Fig. 5g). After the electroforming, a clear reduction of the S atom percentage in the channel region was observed with the atomic ratio of Mo : (S+O) decreased to approximate 1 : 1.2, as shown in Fig. 5e. This observation is consistent with the reduced contrast at the channel region in its corresponding HAADF image (Fig. 5b). The loss of S atoms (forming S vacancies) could result from thermophoresis effect due to the Joule heating: the temperature gradient caused by the increasing current drives the lighter ions away from the channel region (as schematically shown in Figs. 5g, h).

Furthermore, a comparison of *in situ* STEM results between the ON and OFF states could reveal a switching mechanism of GMG devices. It can be seen that there was no noticeable change in the MoS$_{2-x}$O$_x$ layer as shown in the HAADF image (Fig. 5c), suggesting that no intense atomic movement occurred during the reset process. At the same time, the horizontal line-scan exhibits a percentage increase of oxygen atoms in the channel region for the OFF state, with an atomic ratio of Mo: (S+O) changed to be around 1 : 2 (Fig. 5f). The amount of S in the channel region remains the same or even slightly decreases from the ON state to the OFF state. The experimentally observed increase of oxygen atoms in the channel region indicates that the oxygen ions near the channel region are driven (mainly by thermal dissolution effect) towards the channel to fill the sulfur vacancies, and consequently switch the device to the OFF state (as schematically shown in Fig. 5i). Here the filled oxygen ions are more mobile than the sulfur ions likely due to the lower barrier energy for motion. For the set process, since thermophoresis effect[64,65] would dominate due to the steep radial temperature gradient produced by Joule heating, the oxygen ions are driven out of the channel region. The switching mechanism primarily based on the migration of oxygen ions with minor structure change of the channel region is believed to be a major origin of the observed high switching performance.[26]

Based on our results obtained in the *in situ* STEM experiments, both the graphene electrodes and the $MoS_{2-x}O_x$ switching layer, together with their atomically sharp interfaces, are considered to be responsible for the high thermal stability of the GMG devices. Firstly, we found that the channel region still maintains layered crystal structure after the electrical operations (see the HRTEM image of the *in situ* planar GMG device in Supplementary Fig. 7b). By considering the aforementioned extra-high thermal stability of the $MoS_{2-x}O_x$ layer, at elevated temperatures, this could effectively avoid the undesirable migration of oxygen and sulfur ions, and prevent possible reactions between the conduction channel and surrounding regions. Secondly, with a high thermal and chemical stability[44] and good impermeability[66], the graphene electrodes act as protector of the conduction channel at elevated temperatures. Indeed, we noticed that the graphene electrodes maintain excellent crystal structure after switching (Supplementary Fig. 7b). The carbon EDS mapping of the channel region in pristine, ON and OFF states (Supplementary Fig. 8) further demonstrates the compositional stability of the graphene electrodes. With the help of the atomically sharp interface between graphene and $MoS_{2-x}O_x$ layer, the graphene electrodes well confine a stable conduction channel and effectively prevent the migration of the active ions[57] into electrode materials, which is strongly correlated to device failure, especially at elevated temperatures. This also explains in ambient conditions (with the existence of oxygen and humidity), the as-prepared MoS$_2$ membrane with existing sulfur vacancies[67] can be oxidized at quite low temperature (160°C) to be MoS$_{2-x}$O$_x$, which still maintains good layered crystal structure (Fig. 1d). Nevertheless, after being isolated from oxygen and humidity (either in HRTEM vacuum or being well-protected by graphene electrodes), the MoS$_{2-x}$O$_x$ membrane has shown good structure stability at temperatures up to 800°C in HRTEM (Fig. 4) and good electrical switching behaviors at temperatures up to 340°C (Fig. 3).

**Flexible electronic applications.**

Both graphene and MoS$_2$ have shown outstanding mechanical flexibility[45,68], which is ideal for harsh electronics against mechanical stress and other flexible electronic applications[53,69-71]. As a demonstration, we fabricated flexible GMG crossbar structures on PI substrate (Fig. 6a, 6b). The flexible GMG devices could be set at about 0.4 V with a current compliance of 1 mA, and reset at about -0.5 V (with switching curves of a typical device shown in Fig. 6c), which is comparable to the GMG devices on a SiO$_2$/Si substrate. As shown in Fig. 6c, the device exhibits excellent mechanical durability by maintaining critical resistance states and remaining functional over 1000 bending cycles (corresponding to a strain of ~0.6%). Usually it is very challenging for electronic devices to possess both good flexibility and high thermal stability; inorganic material based devices usually lack mechanical flexibility while organic material based devices usually lack good thermal stability.

**Conclusions**

In summary, we fabricated robust memristors based on fully layered 2D materials (graphene/MoS$_{2-x}$O$_x$/graphene), which exhibit repeatable bipolar resistive switching and high thermal stability up to 340°C. We found the atomically sharp interface formed between graphene and MoS$_{2-x}$O$_x$ is responsible for the observed high switching performance. MoS$_{2-x}$O$_x$ layer was found to induce the observed high thermal stability after performing high temperature *in situ* HRTEM studies. Further *in situ* STEM investigations on a cross-sectional device revealed a well-confined conduction channel and a switching mechanism based on the migration of oxygen ions. The atomic layered structure of both switching layer (MoS$_{2-x}$O$_x$) and electrodes (graphene), together with their atomically sharp interfaces, were found to play a crucial role in determining the robustness of the devices. Finally, the mechanical flexibility of such structured devices was demonstrated by showing a good endurance against mechanical bending of over 1000 times, suggesting possible flexible electronic applications. Our realization of robust memristors based on fully 2D materials provides an avenue for future electronics engineering by using vdW heterostructures.

**Methods**

**Device fabrication.** The multi-layer graphene and $MoS_2$ membranes were obtained by using mechanical exfoliation method on 300-nm thick $SiO_2$ wafers, where commercial graphite and $MoS_2$ flakes were used as received (graphite from Kish Graphite; $MoS_2$ from SPI Supplies). The graphene ribbons were defined by standard e-beam lithography (EBL) method, followed by dry etching in an Inductively Coupled Plasma (ICP) system, where $O_2$ was used as etching gas. The thickness of graphene and $MoS_2$ membranes was identified by an atomic force microscopy (AFM). The oxidation of the $MoS_2$ membranes was performed in ambient air on a hot plate at 160°C for 1.5 h, with subsequent rapid cooling. The graphene/$MoS_{2-x}O_x$/graphene (GMG) structure was stacked by using standard polyvinyl alcohol (PVA) transfer method[38]. The metal conductive layer (5 nm Ti/50 nm Au) was deposited through standard E-beam deposition process. To fabricate Au/$MoS_{2-x}O_x$/Au (AMA) devices, we first deposited an Au bottom electrode (40nm thick, 1 μm wide) on a 300-nm thick $SiO_2$ wafer through standard EBL and E-beam deposition processes. A $MoS_{2-x}O_x$ membrane was then transferred onto the bottom electrode. An Au top electrode (40 nm thick, 1 μm wide) was finally deposited onto the $MoS_{2-x}O_x$ membrane perpendicular to the bottom electrode.

**Characterizations.** Current–voltage switching curves and resistance measurements were performed using an Agilent B1500A parameter analyzer. The high temperature electrical measurements of GMG and AMA devices were performed on a hot plate in ambient air. The temperatures of the devices were further confirmed by a Fluke non-contact infrared thermometer.

**High temperature *in situ* HRTEM experiments.** $MoS_2$ nanosheets were prepared by ultrasonic exfoliation using commercial $MoS_2$ bulk samples (SPI Supplies). The dispersion was then dropped onto a DENSsolution *in situ* heating chip. Before the *in situ* HRTEM experiments, sample was heated to 160°C in ambient air and kept 1.5 h for thermal oxidation. HRTEM images were acquired by FEI Tecnai F20 at 200 KV. DENSsolution DH30 system was used for *in situ* heating experiments.

***In situ* Cross-section STEM experiments.** High quality Pt (conductive protection layer)/graphene/$MoS_{2-x}O_x$/graphene/Si (conductive silicon substrate) samples were manufactured on our home-made *in-situ* electrical chip (designed for DENSsolution DH30 holder) by a Helios 600i dual-beam FIB system. The cross-sectional lamellae were thinned down to approximate 100 nm at an accelerating voltage of 30 kV with 0.79 nA, followed by two steps of fine polish at 5 kV accelerating voltage with 0.12 nA and 2 kV accelerating voltage with a small current of 68 pA, respectively. The lamellae were fixed on two electrodes of the *in situ* electrical chip by Pt deposition in the FIB system. The STEM and HRTEM images were obtained on a FEI Titan Cubed G2 60-300 aberration corrected S/TEM. The operation voltage of 60 kV was used to reduce electron beam damage to graphene and $MoS_{2-x}O_x$. EDS analyses were carried out using Bruker SuperEDX four-detector system. Real-time electrical measurements of the *in situ* samples were performed using a Keithley 2636A dual channel digital source meter unit connected with a DENSsolution double-tilt heating and biasing holder. In this experiment, thicker graphene membranes (>50 nm) were used for both top and bottom electrodes to protect the $MoS_{2-x}O_x$ layer from being damaged by the high energy focused ion beam (Gallium ion), and to avoid the disturbance of the Si and Pt atoms during the switching process.

of human skin. *Nat. Mater.* **12**, 938-944 (2013).

## Acknowledgements

This work was supported in part by the National Key Basic Research Program of China (2015CB921600, 2015CB654901 and 2013CBA01603), National Natural Science Foundation of China (61625402, 61574076, 11474147 and 11374142), Natural Science Foundation of Jiangsu Province (BK20140017 and BK20150055), Fundamental Research Funds for the Central Universities, and Collaborative Innovation Center of Advanced Microstructures.

## Author contributions

M. W. and S. C. contributed equally to this work. F. M. and M. W. conceived the project and designed the experiments. M. W., C. P., C. W., X. L. and K. X. performed the device fabrication and electrical measurements. S. C., M. W. and P. W. carried out the *in situ* HRTEM and cross-section STEM experiments and analyses. M. W., F. M., X. L. and J. J. Yang conducted the data analyses and interpretations. F. M., M. W., S. C., P. W. and J. J. Yang co-wrote the paper, and all authors contributed to the discussions and preparation of the manuscript.

## Competing financial interests

The authors declare no competing interests.

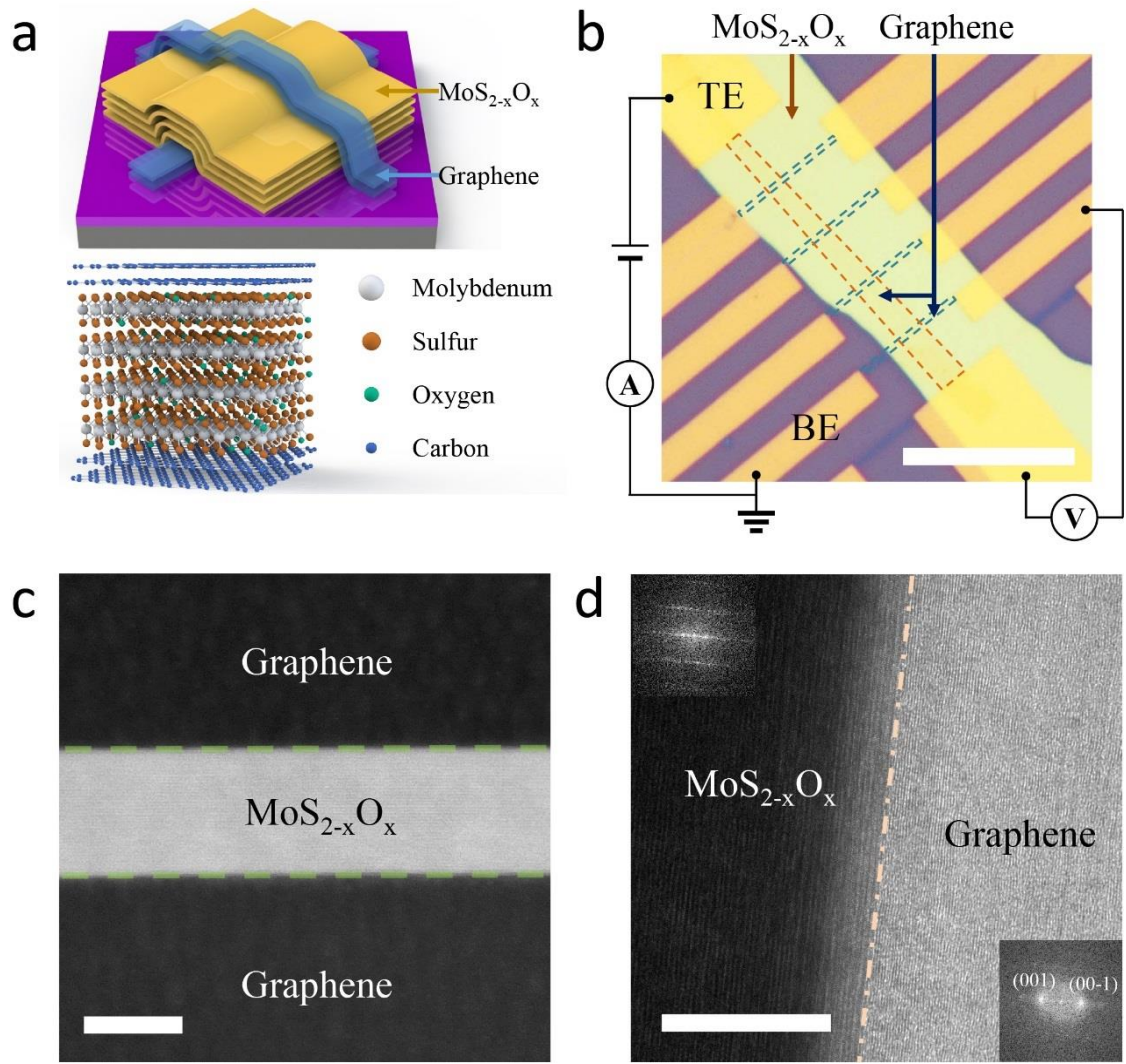

**Figure 1 | GMG devices and cross-section images.** (a) Top: schematic drawing of the GMG devices. Bottom: crystal structure of the GMG devices. (b) Optical microscope image and measurement setup of a GMG device. Scale bar: 25 μm. The top and bottom graphene electrodes are highlighted by orange and blue dashed boxes, respectively. Four-probe measurements were carried out to rule out the resistance of graphene electrodes. (c) Cross-section HAADF image of a pristine GMG device. Scale bar: 20 nm. (d) Cross-section HRTEM image of a pristine GMG device. Top and bottom insets: power spectra of the $MoS_{2-x}O_x$ layer and graphene in the HRTEM image, respectively. $MoS_{2-x}O_x$ maintains layered crystal structure after the oxidization, with excellent contact with graphene. Scale bar: 20 nm.

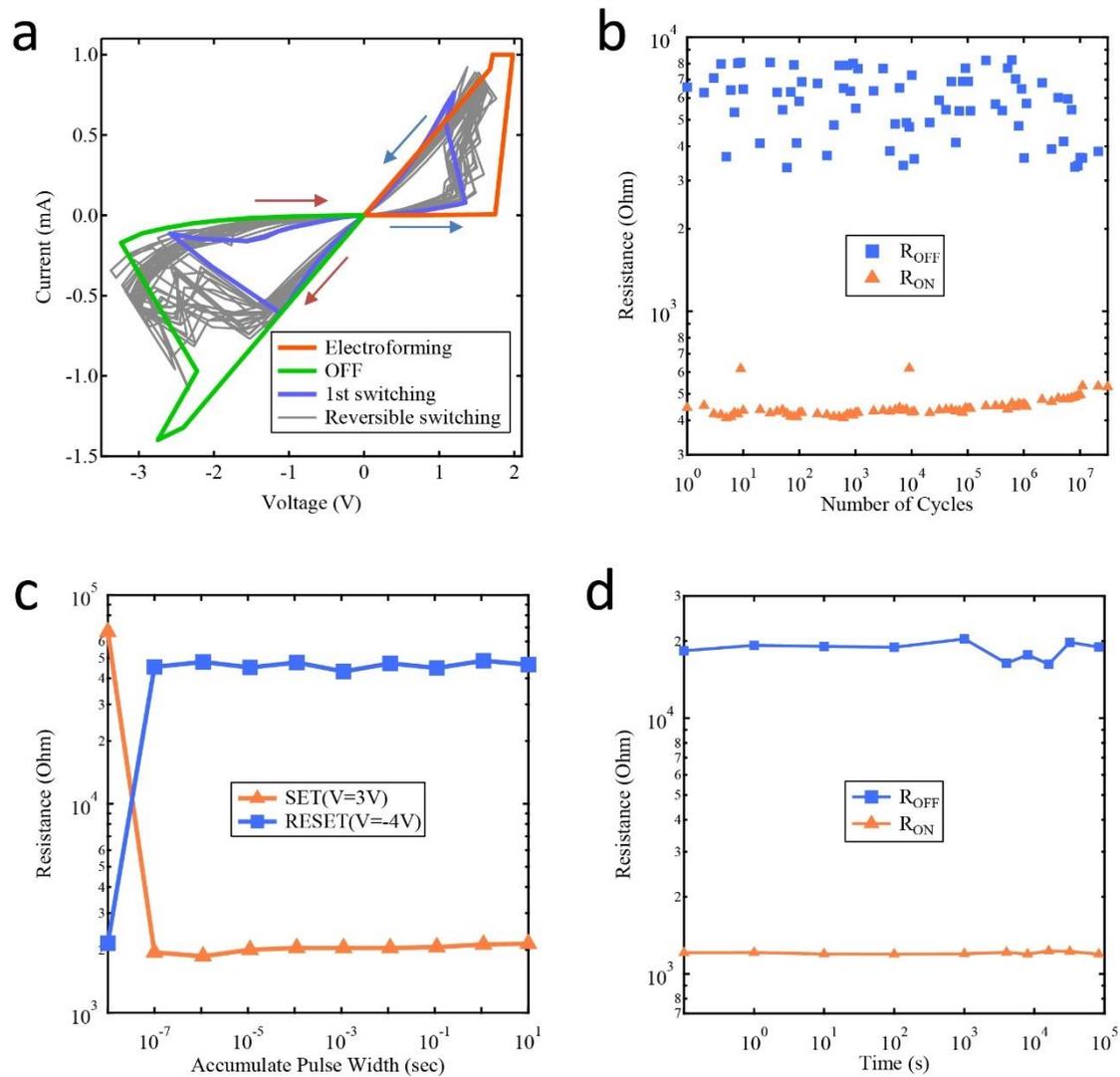

**Figure 2 | Electrical characterizations of the GMG devices.** (a) DC switching curves of a GMG device with an electroforming current compliance of 1 mA. The green line is the reset process after electroforming step. The arrows indicate the switching directions. (b) Over $2 \times 10^7$ switching cycles obtained with 1μs width pulses applied (about +3.5 V and -4.8 V dropped across the device for set and reset, respectively). (c) Switching speed test of a GMG device. The voltage drop across the device is about +3 V and -4 V for set and reset, respectively. The device can be switched in less than 100ns and keep its resistance state with wider pulses applied. (d) About $10^5$ s retention time measured in a GMG device. The blue and orange dots represent OFF and ON state resistances, respectively. The resistance values were read at $V_r = 0.1$ V.

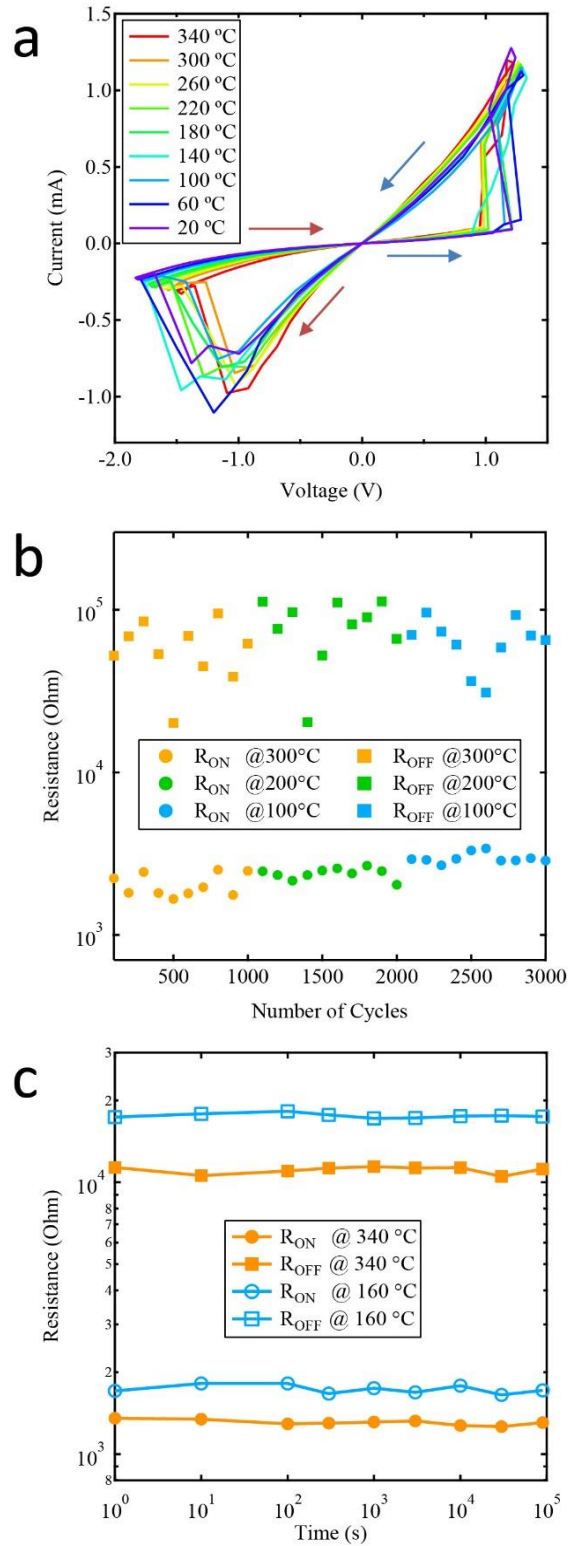

**Figure 3 | Electrical characterizations of the GMG devices at elevated temperatures.** (a) Switching curves of a GMG device at different temperatures. The arrows indicate the switching directions. The GMG device was able to operate at a high temperature up to 340°C, demonstrating high thermal stability. (b) 1000-time endurance test under 1 μs pulse at 300°C (orange), 200°C (green) and 100°C (blue), respectively.

(c) Retention time at 340°C (orange) and 160°C (blue). The resistance values were read at $V_r = 0.1$ V. The good endurance and retention time further demonstrate the high thermal stability of the GMG devices.

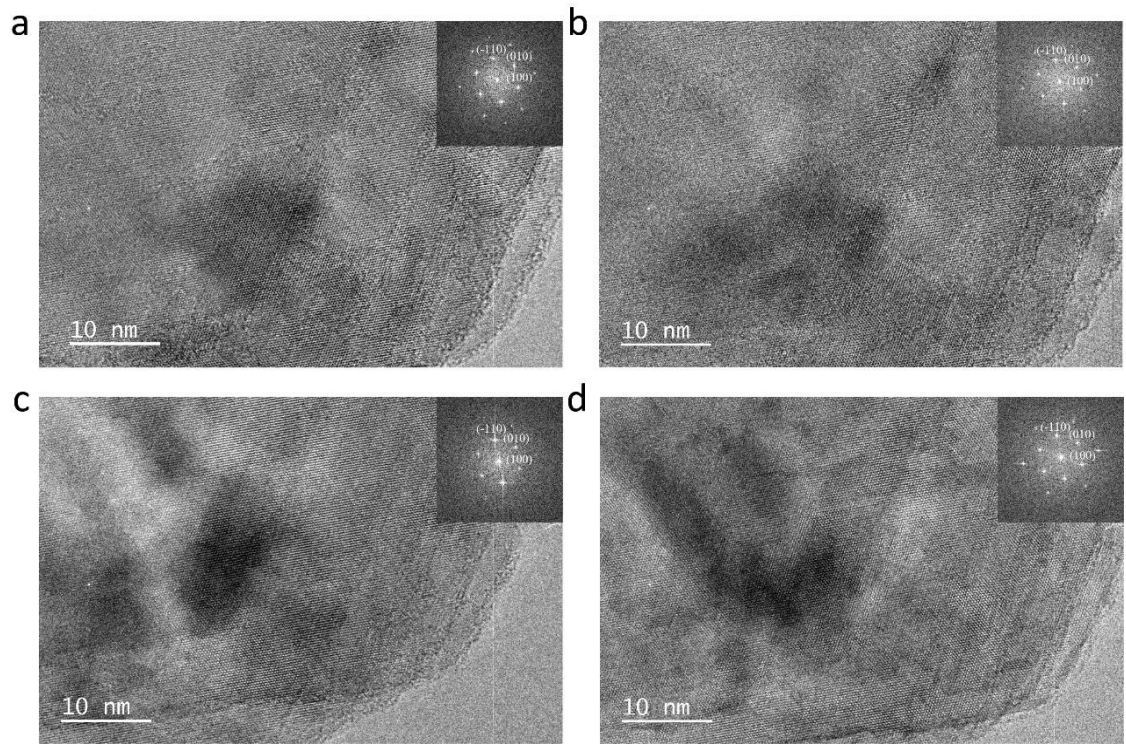

**Figure 4 | *In situ* HRTEM observation of MoS$_{2-x}$O$_x$ at elevated temperatures.** (a), (b), (c) and (d) are HRTEM images of the same MoS$_{2-x}$O$_x$ membrane at room temperature, 300°C, 600°C and 800°C, respectively. MoS$_{2-x}$O$_x$ maintains excellent crystal structure at 800°C. Inserts: power spectra of the corresponding HRTEM images of the MoS$_{2-x}$O$_x$ membrane at different temperatures, all of which show the reflections corresponding to the MoS$_{2-x}$O$_x$ structure. The lattice constant of the MoS$_{2-x}$O$_x$ membrane is 3.16 Å at 20°C and 3.19 Å at 800°C as calculated from the power spectrum images. All images were taken after the target temperatures were kept stable for 10 min.

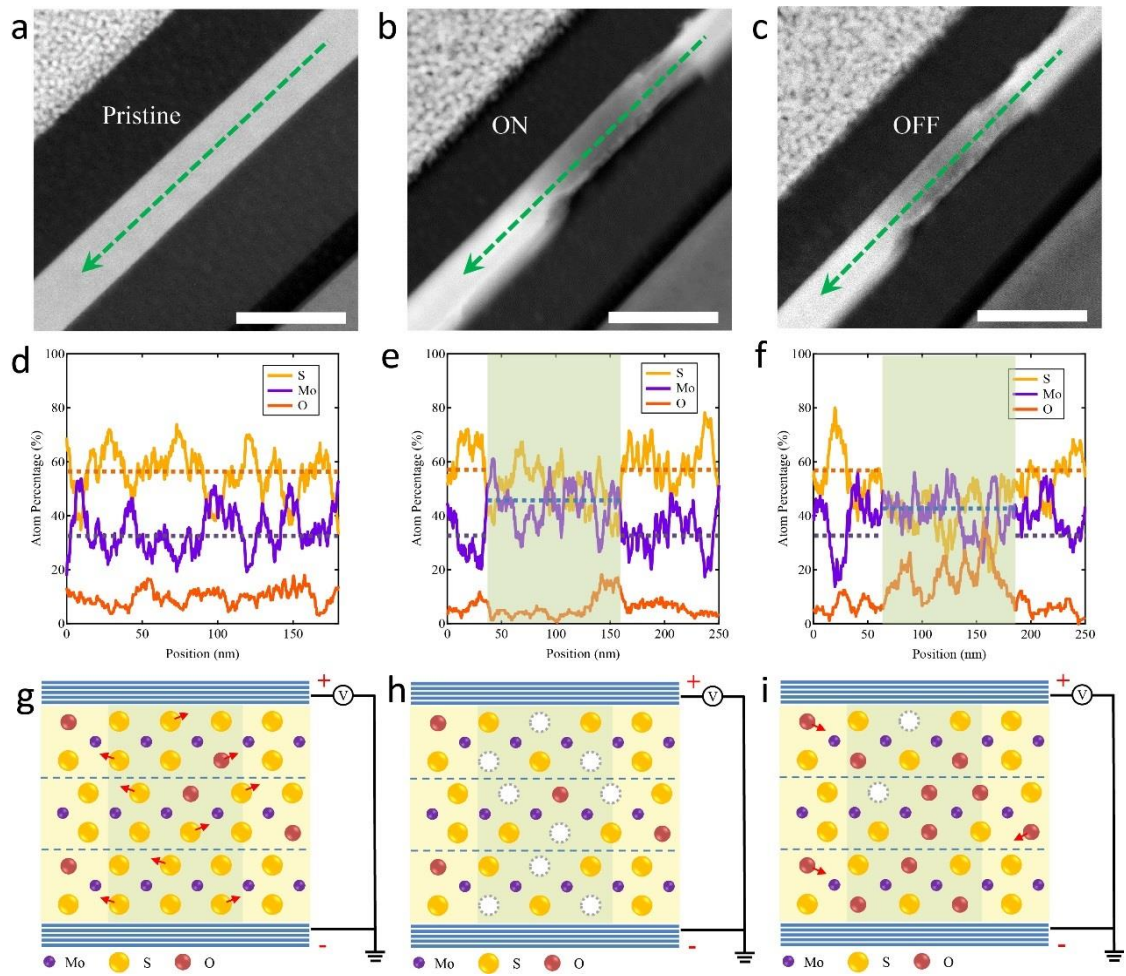

**Figure 5 |** *In situ* **STEM observation of the conduction channel in GMG devices.** (a), (b), (c) Cross-section HAADF image of a single GMG device in the (a) pristine state, (b) ON state and (c) OFF state, respectively. A dark region in the $MoS_{2-x}O_x$ layer appears in (b) ON state and remains unchanged when the device is switched to (c) OFF state. All scale bars: 50 nm. (d), (e), (f) EDS line-scan profiles of the green arrows in (a), (b), (c), respectively. The numbers of Mo, S and O were measured and normalized as atom percentage. (d) $MoS_{2-x}O_x$ layer in pristine state shows a uniform atom distribution, where Mo : (S+O) ≈ 1 : 2. (e) A reduction of S atoms was observed in the dark region, where Mo : (S+O) is approximately 1 : 1.2. (f) An increase of O atoms was observed in the conduction channel for OFF state, yielding a ratio of Mo : (S+O) back to the value near 1 : 2. (g), (h), (i) Schematic structure diagrams illustrating resistance-switching mechanism in the GMG devices. (g) Electroforming (from pristine state) step, showing a uniform elemental distribution in pristine state and the appearance of sulfur vacancies during electroforming. (h) ON state with sulfur vacancies formed within the channel region. (i) OFF state showing that the sulfur vacancies are filled with oxygen ions from the regions surrounding the channel.

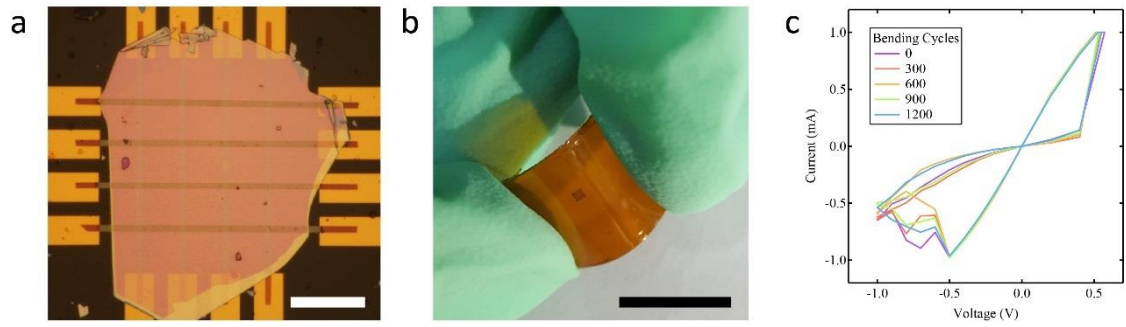

**Figure 6 | Flexible GMG devices.** (a) Optical microscope image of a GMG crossbar device on a PI substrate. Scale bar: 20 μm. (b) A photograph of a crossbar array under bending. Scale bar: 10 mm. (c) Switching curves of a GMG device against repeated mechanical bending to a 1-cm radius curvature.

**Supplementary Figures**

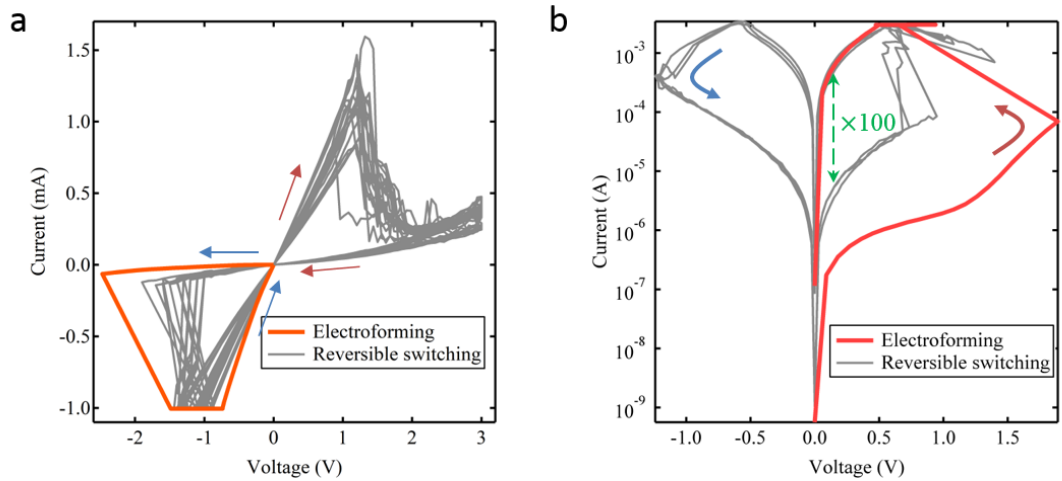

**Supplementary Figure 1 | Additional DC switching curves of GMG devices.** (a) DC switching curves of a GMG device with a negative electroforming voltage and the current compliance of 1 mA. (b) DC switching curves of a GMG device with a current compliance of 3 mA for electroforming and set process. The ON/OFF ratio of the reversible switching reaches about 100. The arrows indicate the switching directions.

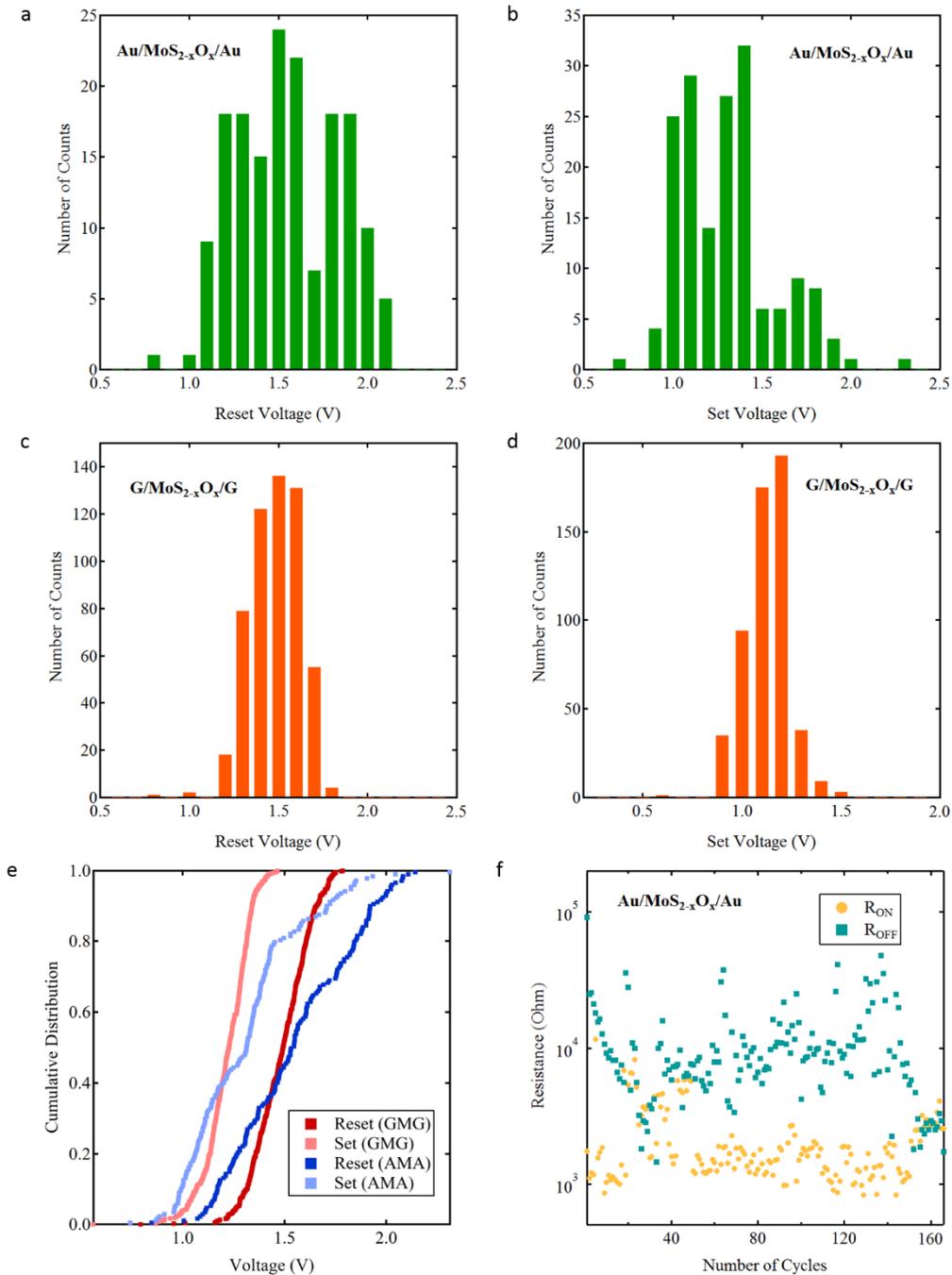

**Supplementary Figure 2 | The statistics of the switching parameters in AMA and GMG devices.** (a), (b) Distribution of (a) reset voltage (absolute value) and (b) set voltage based on 550 switching cycles in AMA devices. (c), (d) Distribution of (c) reset voltage (absolute value) and (d) set voltage based on 150 switching cycles in GMG devices. (e) The cumulative distribution of reset voltage (absolute value) and set voltage in GMG and AMA devices. (f) The endurance test for AMA devices. The resistance values were read at $V_r = 0.1$ V.

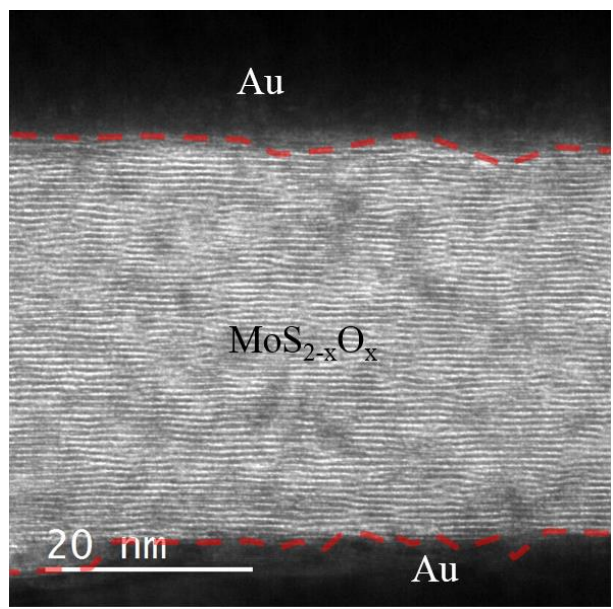

**Supplementary Figure 3 | Cross-section observation of an AMA device.** Cross-section HRTEM image of a typical AMA device. Scale bar: 20 nm.

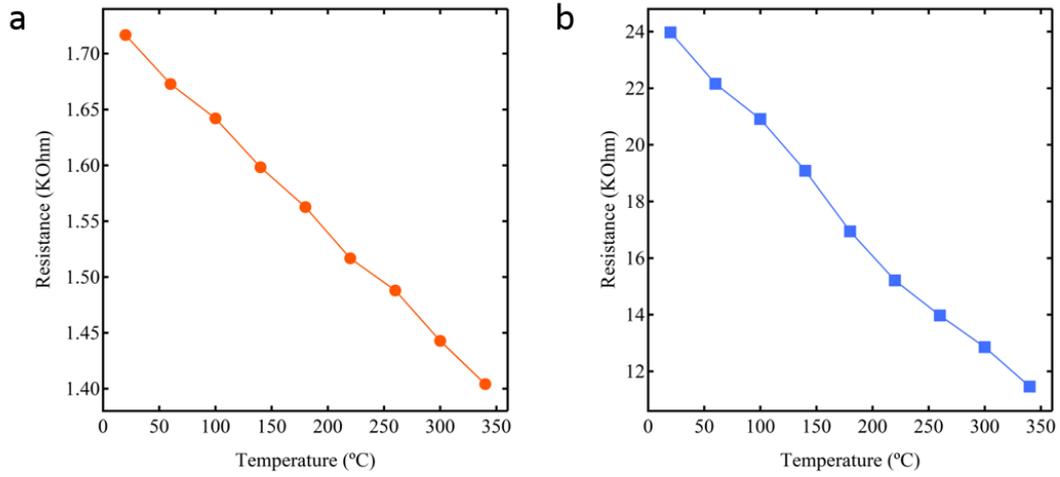

**Supplementary Figure 4 | Temperature dependence of ON/OFF states of GMG devices.** (a) R-T characteristics obtained from the ON state of a GMG device. (b) R-T characteristics obtained from the OFF state of the same device. The resistance values were read at $V_r = 0.1$ V.

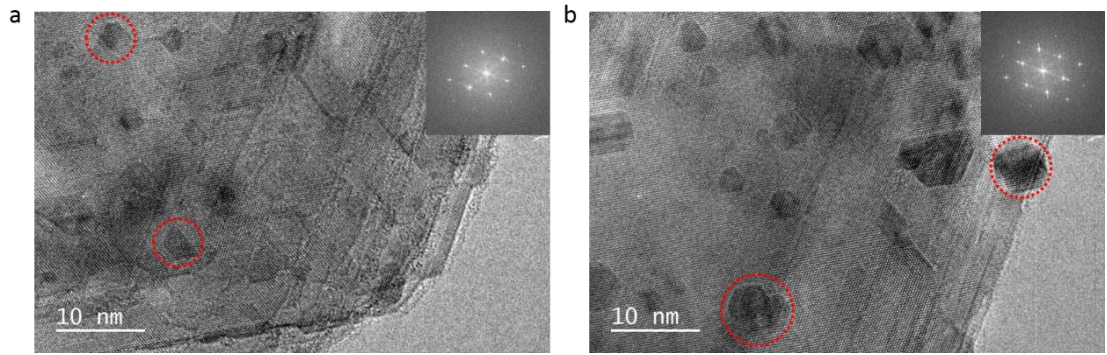

**Supplementary Figure 5 | *In situ* HRTEM observation of MoS$_{2-x}$O$_x$ at 900°C and 1000°C.** HRTEM image of a MoS$_{2-x}$O$_x$ membrane at 900°C (a) and 1000°C (b). Red dashed circles highlight the formed clusters. Inserts: power spectra of the corresponding HRTEM image of the MoS$_{2-x}$O$_x$ membrane. All images were taken after the target temperatures were kept stable for 10 min.

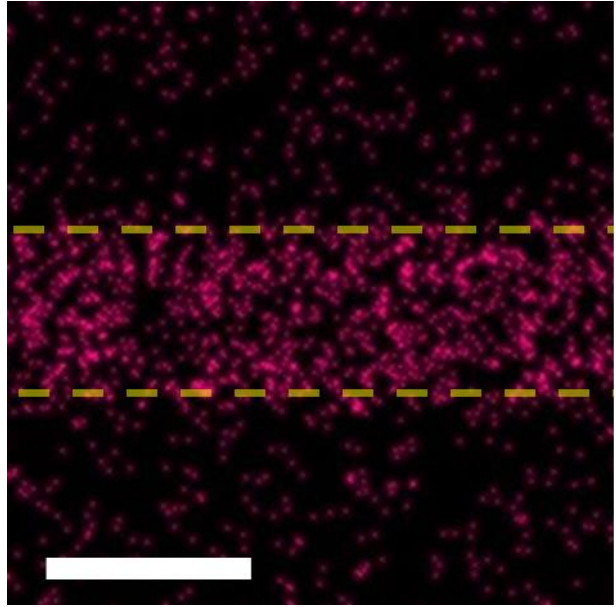

**Supplementary Figure 6 | Cross-section Oxygen EDS mapping of a GMG device at pristine state.** It shows a uniform oxygen atom distribution. Scale bar: 25nm.

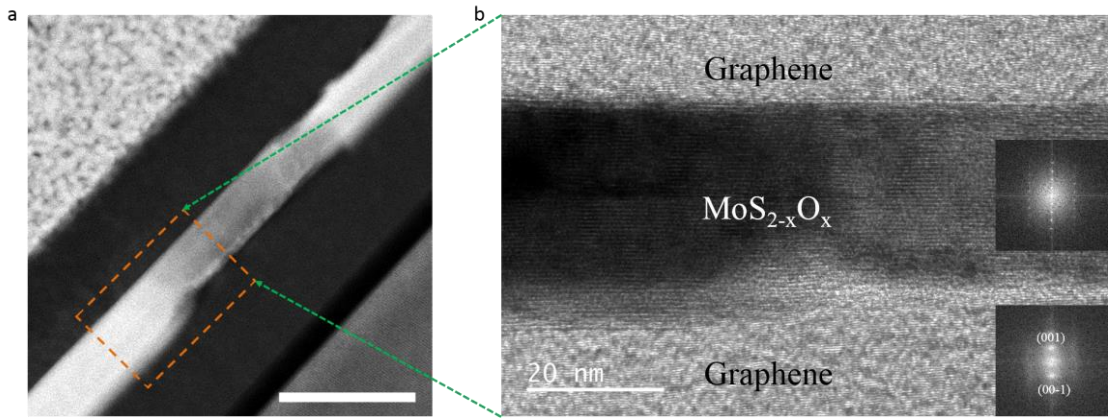

**Supplementary Figure 7 | Cross-section HAADF and HRTEM images of a GMG device in OFF state.** (a) Cross-section HAADF image of the GMG device in OFF state, which is the same as the one shown in Fig. 5c. Scale bar: 50 nm. (b) HRTEM image taken from the area indicated with the dashed rectangle in (a), showing a sharp interface and layered crystal structure of both graphene and $MoS_{2-x}O_x$ layers after the switching process. Middle and bottom insets: power spectra of the $MoS_{2-x}O_x$ layer and graphene in the HRTEM image, respectively.

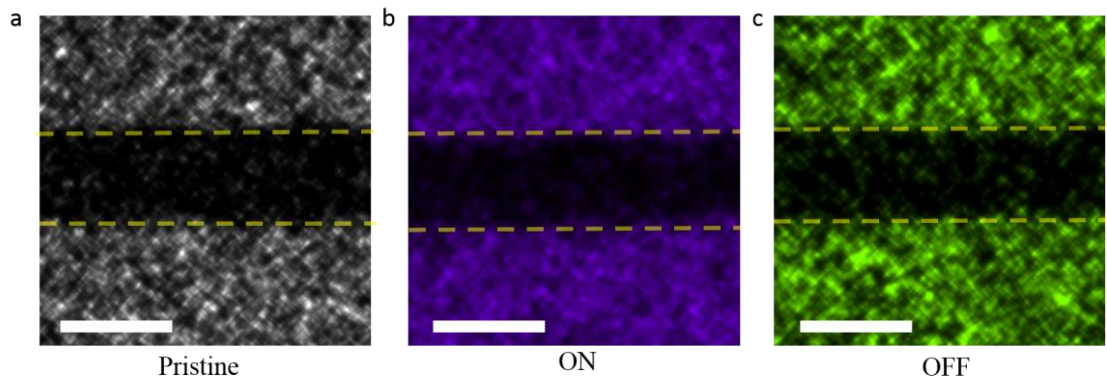

**Supplementary Figure 8 | Cross-section Carbon EDS mappings of the GMG device in the *in situ* observation experiment.** Carbon EDS mappings of the conduction channel in (a) pristine state, (b) ON state and (c) OFF state, respectively. All scale bars: 25 nm.

# Supplementary Notes

**Supplementary Note 1: The switching polarity in GMG devices**

In GMG devices, the switching polarity is determined by the electroforming voltage polarity. The switching curves with negative electroforming voltage are shown in Supplementary Figure 1a. The switching behaviors are nearly the same as those with positive forming voltage (Fig. 2a). Different from the memristors with asymmetric structure, the electroforming process with either positive or negative forming voltages in GMG devices defines the switching interface, and thus the switching polarity.

**Supplementary Note 2: The statistics of the switching parameters in AMA and GMG devices**

To investigate the difference between graphene electrode and Au electrode devices, we collated the statistics of the switching parameters in AMA and GMG devices. Compared with the AMA devices, the GMG devices have shown more concentrated distributions of both reset and set voltages (as shown in Supplementary Figures 2a-d), suggesting a more stable switching and better cycle-to-cycle variability in the GMG devices. The better cycle-to-cycle variability of the GMG devices is further demonstrated by the statistics shown in Supplementary Figure 2e: the resistances of ON-state and OFF-state in GMG devices are more stable than AMA devices; the cumulative distribution of reset and set voltages in GMG device is also more concentrated and smoother than AMA devices.

**Supplementary Note 3: HRTEM observation of an AMA device**

As a comparison to GMG devices, a cross-section sample of AMA device was fabricated. HRTEM image of the AMA cross-section sample is shown in Supplementary Figure 3, which exhibits rougher interfaces (highlighted by red dashed lines) with overlap areas between Au electrodes and $MoS_{2-x}O_x$ layer.

**Supplementary Note 4: *In situ* HRTEM observation of $MoS_{2-x}O_x$ at 900°C and 1000°C**

To further examine the thermal stability of $MoS_{2-x}O_x$, we also took the HRTEM images of $MoS_{2-x}O_x$ at 900°C and 1000°C. As shown in Supplementary Figure 5a, the atoms begin to migrate and form clusters (red dashed circles) at 900°C. New spots appear on FFT images (insets), suggesting a structure change occurs in $MoS_{2-x}O_x$. When the temperature rises to 1000°C, the migration of atoms become more intense and the clusters (red dashed circles in Supplementary Figure 5b) become bigger. All HRTEM images were taken after the target temperatures were kept stable for 10 min.